\documentclass{ckm}                 

\usepackage{mathptmx}          
\usepackage{amsmath} 
\usepackage{citesort}
\usepackage{epsfig}

\newcommand{\beqa}{\begin{eqnarray}}
\newcommand{\eeqa}{\end{eqnarray}}
\newcommand{\beq}{\begin{equation}}
\newcommand{\eeq}{\end{equation}}
\newcommand{\bal}{\begin{align}}
\newcommand{\eal}{\end{align}}
\renewcommand{\Re}{{\sl Re}}

\def\gsim{\ \rlap{\raise 3pt \hbox{$>$}}{\lower 3pt \hbox{$\sim$}}\ }
\def\lsim{\ \rlap{\raise 3pt \hbox{$<$}}{\lower 3pt \hbox{$\sim$}}\ }

\def\Dbar{\overline D}
\def\DbarZ{\overline{D}^0}

\confname{Workshop on the CKM Unitarity Triangle, IPPP Durham, April
  2003}

\title{Determination of  the angle $\gamma$ using multibody $D$ decays in $B^\pm \to D K^\pm$}

\author{Anjan Giri\addressmark{a}, Yuval Grossman\addressmark{a}, Abner Soffer\addressmark{b} and Jure
Zupan \addressmark{a,c}\thanks{Talk given by J. Zupan, based on \cite{Giri:2003ty}.}}

\address[a]{Department of Physics,     
Technion--Israel Institute of Technology,
Technion City, 32000 Haifa, Israel}
\address[b]{Department of Physics, Colorado State University, 
Fort Collins, CO 80523}
\address[c]{J.~Stefan Institute, Jamova 39, P.O. Box 3000,1001
Ljubljana, Slovenia}

\begin{document}

\begin{abstract}We describe  a method for determining $\gamma$ using $B^\pm\to D K^\pm$
decays followed by a multibody $D$ decay. In the talk we focus on  $
K_S\,\pi^-\pi^+$ final state, but other modes such as $D \to K_S\,K^-K^+$ and $D \to K_S\,
\pi^-\pi^+\pi^0$ can also be used.  The main advantages of the method are that it uses
only Cabibbo allowed $D$ decays, and that large strong phases are
expected due to the presence of resonances. Since no
knowledge about the resonance structure is needed, $\gamma$ can be
extracted without any hadronic uncertainty.
\end{abstract}

\maketitle


\section{Basic idea}
\begin{figure}[h!]
\begin{center}
\includegraphics[width=3.3cm]{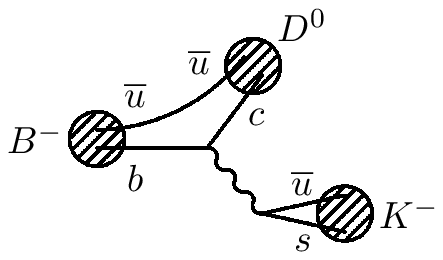}\hspace{0.2cm}\raisebox{1.2cm}{$\Longleftrightarrow$}\raisebox{0.5cm}{\includegraphics[width=3.3cm]{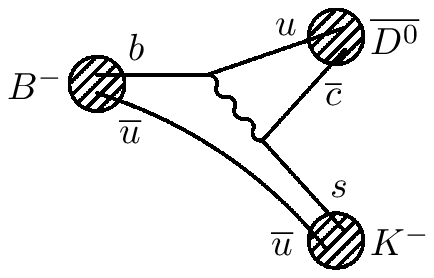}}
\vspace{-0.6cm}
\caption{\footnotesize{The dependence on $\gamma$ arises from the interference between the $B^\pm\to D^0 K^\pm$ and $B^\pm\to \DbarZ
K^\pm$ decay amplitudes. }} \label{fig-interf}
\end{center}
\end{figure}

The theoretically cleanest way of determining the angle 
\begin{equation} 
\gamma=\arg(-V_{ud}V^*_{ub}/V_{cd}V_{cb}^*),
\end{equation}
is to utilize the interference between the $b\to c\bar{u}s$ and $b\to
u\bar{c} s$ decay amplitudes
\cite{Gronau:1991dp,Atwood:1996ci,Atwood:2000,Soffer:1999dz,other,Aleksan:2002mh,Gronau:2002mu},
see Fig. \ref{fig-interf}.
The salient feature of these transitions is that they involve only
distinct quark flavors and therefore do not receive any  penguin contributions. In the original idea by
Gronau and Wyler (GW) \cite{Gronau:1991dp} the $B^\pm\to D_{CP} K^\pm$
decay modes are used, where $D_{CP}$ represents a $D$ meson which
decays into a CP eigenstate.  The dependence on $\gamma$ arises from
the interference between the $B^\pm\to D^0 K^\pm$ and $B^\pm\to \DbarZ
K^\pm$ decay amplitudes. The main advantage of the GW method is that,
in principle, the hadronic parameters can be cleanly extracted from
data, by measuring the $B^\pm\to D^0 K^\pm$ and $B^\pm\to \DbarZ
K^\pm$ decay rates.

\vspace{-1cm}
\vspace{-0.8cm}

In practice, however, measuring $\gamma$ in this way is not an easy
task.  Due to the values of the CKM coefficients and color
suppression, the ratio between the two interfering amplitudes, $r_B$
[see Eq. (\ref{weakphase})], is expected to be small, of order
$10\%-20\%$. This reduces the sensitivity to $\gamma$, which is
roughly proportional to the magnitude of the smaller amplitude. It
also leads to experimental difficulties in measuring the
color suppressed  $B^-\to \overline{D^0} K^-$ mode (and its charge
conjugate) preventing a straightforward application of the  GW method \cite{Atwood:1996ci}.
There exist a number of extensions of the original GW proposal which avoid
the problem by not relying on the measurement of $B^-\to
\overline{D^0} K^-$ amplitude \cite{Atwood:1996ci,Atwood:2000,other}. Instead several
different decay modes of $D$ mesons such as quasi
two-body $D$ decays with one particle a resonance (e.g. $D^0\to
K^{*+}\pi^-$ \cite{Atwood:1996ci,Atwood:2000}) are used. Since these are
really three body decays (for instance in the example mentioned, $
K^{*+}$ decays strongly to $K^0 \pi^+$ or $K^+\pi^0$) one can pose
the following questions : 
\begin{itemize}
\item
Can one use the complete phase space of such three-body $D$ decays for $\gamma$ extraction? 
\item
Is it possible to avoid fits to Breit-Wigner forms in doing the Dalitz
plot analysis?
\end{itemize}

As we show in the following, the answer to both of these questions is positive.
The first question was raised already in \cite{Atwood:2000}, however,
most of the results  and applications  we present are
new.  For the sake of concreteness, we
concentrate on the $D \to K_S\,\pi^-\pi^+$ decay mode, while an extension
to a larger set of the decay modes can be found in \cite{Giri:2003ty}.  The advantage
of using the chosen decay chain is threefold. First, one expects large
strong phases due to the presence of resonances. Second, only Cabibbo
allowed $D$ decay modes are needed. Third, the final state involves
only charged particles, which have a higher reconstruction efficiency
and lower background than neutrals.  The price one has to pay is that
a Dalitz plot analysis of the data is needed.  We describe how to do
the Dalitz plot analysis in a model-independent way, and explore the
advantages gained by introducing verifiable model-dependence.  The
final balance between the advantages and disadvantages depends on
yet-to-be-determined hadronic parameters and experimental
considerations. Finally, we mention that an equivalent formalism to the one we present
below has been  independently  developed by Atwood and Soni in \cite{Atwood:2003mj}.

\section{Model independent determination of $\gamma$}\label{model-indep}
Let us focus on the following cascade decay \footnote{In the following discussion we neglect 
$D^0-\bar{D^0}$ mixing, which is a good approximation in the context
of the Standard Model, for more details see app. A of \cite{Giri:2003ty}.}
\begin{equation}
B^- \to D K^- \to (K_S \pi^- \pi^+)_D K^-,
\end{equation}
and  define the amplitudes
\begin{align}
A(B^- \to D^0 K^-)&\equiv A_B \label{AB},\\ 
A(B^- \to {\DbarZ} K^-) &\equiv
A_B r_B e^{i(\delta_B-\gamma)}.\label{weakphase}
\end{align}
The same definitions apply to the amplitudes for the CP conjugate
cascade $B^+ \to D K^+ \to (K_S \,\pi^+\pi^-)_D K^+$, with the change
of weak phase sign $\gamma\to -\gamma$ in (\ref{weakphase}). We have set the strong phase of $A_B$ to zero by convention, so that
$\delta_B$ is the difference of strong phases between the two
amplitudes.  For the CKM elements, the usual convention of the weak
phases has been used. 
The value of $|A_B|$ is known from the measurement of
the $B^-\to D^0 K^-$ decay width using flavor specific decays of $D^0$.
The amplitude $A(B^- \to \DbarZ K^-)$ is color
suppressed and cannot be determined from experiment in this way 
\cite{Atwood:1996ci}. The color suppression together
with the experimental values of the ratio of the relevant CKM
elements leads to the theoretical expectation $r_B\sim 0.1-0.2$ (see
recent discussion in \cite{Gronau:2002mu}).

For the three-body $D$ meson decay we define
\begin{equation}
\begin{split}
\label{CP-for-D}
A_D(s_{12},s_{13}) &\equiv A_{12,13}\,e^{i\delta_{12,13}}\\
&\equiv  A(D^0 \to K_S(p_1) \pi^-(p_2) \pi^+(p_3))\\
& =A(\DbarZ \to K_S(p_1) \pi^+(p_2) \pi^-(p_3)),
\end{split}
\end{equation}
where $s_{ij}=(p_i+p_j)^2$, and $p_1,p_2,p_3$ are the momenta of the
$K_S, \pi^-,\pi^+$ respectively. We also set the magnitude
$A_{12,13}\ge0$, such that $\delta_{12,13}$ can vary between $0$ and
$2\pi$.  In the last equality the CP symmetry of the strong
interaction together with the fact that the final state is a spin zero
state has been used. With the above definitions, the amplitude for
the cascade decay is
\begin{equation}
\begin{split}
A(B^-&\to (K_S \pi^- \pi^+)_D K^-)=\\
A_B& {\cal P}_D
\big(A_D(s_{12},s_{13})
+
r_B e^{i(\delta_B-\gamma)}A_D(s_{13},s_{12})\big), \label{amplitude}
\end{split}
\end{equation}
where  ${\cal P}_D$ is the $D$ meson propagator.  Next, we write down the
expression for the reduced partial decay width
\begin{equation}
\begin{split}
 d\hat\Gamma&(B^- \to (K_S \pi^-\pi^+)_D K^-)=
 \Big(A_{12,13}^2 +r_B^2 \, A_{13,12}^2\\ 
&+ 2 r_B\;
\Re\negthickspace\left[A_D(s_{12},s_{13})\,A_D^*(s_{13},s_{12})\,e^{-i(\delta_B-\gamma)}
\right]\Big) dp ,\label{decay-width}
\end{split}
\end{equation}
where $dp$ denotes the phase space variables, and we used 
the extremely accurate  narrow width
approximation for the $D$ meson propagator.

The moduli of the $D$ decay amplitude $A_{12,13}$ can be measured
from the Dalitz plot of the $D^0\to K_S\pi^-\pi^+$ decay. To perform
this measurement the flavor of the decaying neutral $D$ meson has to
be tagged. This can be best achieved by using the charge of the soft
pion in the decay $D^{*+}\to D^0 \pi^+$. 
However, the phase $\delta_{12,13}$ of the $D$ meson decay amplitude 
is not measurable in B-factories without further model dependent
assumptions.  
If the three-body decay $D^0\to K_S\pi^-\pi^+$ is assumed to
be resonance dominated, the Dalitz plot can be fit to a sum of
Breit-Wigner functions, determining also the relative phases of the
resonant amplitudes.  This is further discussed in section
\ref{Breit-Wigner}.  The other option is to use data from charm
factory, where weighted averages of the sine and the cosine of the relevant phase difference may
be measured (see section~\ref{improving}).  Here we assume that no  charm factory data is available
and develop the formalism without any model
dependent assumptions.

Obviously, to compare with the data, an
integration over at least some part of the Dalitz plot has to be
performed. We therefore partition the Dalitz plot into $n$ bins and
define
\begin{subequations}\label{defvar}
\begin{align}
c_i &\equiv \int_i dp\; 
A_{12,13}\,A_{13,12}\,\cos(\delta_{12,13}-\delta_{13,12}),
\label{ci}\\
s_i &\equiv \int_i dp\;  
A_{12,13}\,A_{13,12}\sin(\delta_{12,13}-\delta_{13,12}),
\label{si}\\
T_i &\equiv \int_i dp\;  
A_{12,13}^2, \label{Bi}
\end{align}
\end{subequations}
where the integrals are done over the phase space of the $i$-th
bin. The variables $c_i$ and $s_i$ contain differences of strong phases and
are therefore unknowns in the analysis. The variables $T_i$, on
the other hand, can be measured from the flavor tagged $D$ decays as
discussed above, and are assumed to be known inputs into the analysis.

\begin{figure}
\begin{center}
\epsfig{file=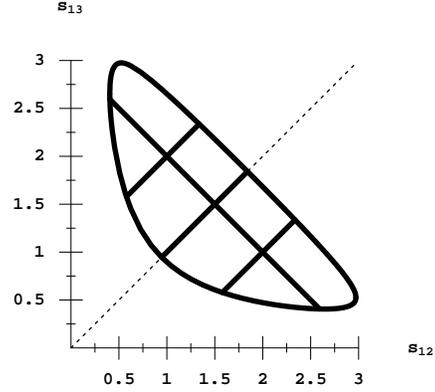, height=5.7cm}
\vspace{-0.5cm}
\caption{\footnotesize{The partitions of Dalitz plot as discussed in
text. The symmetry axis is the dashed line. On the axes we have $s_{12}=m_{K_s\pi^-}^2$ and $s_{13}=m_{K_s\pi^+}^2$ in $\text{GeV}^2$. }} \label{fig1}
\end{center}
\vspace{-0.7cm}
\end{figure}

Due to the symmetry of the interference term, it is convenient to use
pairs of bins that are placed symmetrically 
about the $12 \leftrightarrow 13$ line, as
shown in Fig. \ref{fig1}. Consider an even, $n=2k$, number of bins. The
$k$ bins lying below the symmetry axis are denoted by index $i$,
while the remaining bins  are indexed with $\bar{i}$.
The $\bar{i}$-th bin is obtained by  mirroring the $i$-th bin over
the axis of symmetry. The variables
$c_i, s_i$ of the $i$-th bin are related to the variables of the
$\bar{i}$-th bin by
\begin{equation}
c_{\,\bar{i}}=c_i, \qquad s_{\,\bar{i}}=-s_i, \label{vanish}
\end{equation}
while there is no relation between $T_i$ and $T_{\bar{i}}$.
Note that had one used $12\leftrightarrow 13$ symmetric bins centered
on the symmetry axis, one would have had  $s_i=0$.

Together with the information available from the $B^+$ decay, we 
arrive at a set of $4k$ equations
\begin{subequations}\label{relations4k}
\begin{align}
\begin{split}\label{11a}
\hat\Gamma^-_i &\equiv 
\int_i d\hat\Gamma(B^- \to (K_S \pi^-\pi^+)_D K^-)=\\
&T_i  +r_B^2 T_{\bar{i}}\; + 
2 r_B [\cos(\delta_B-\gamma) c_i + \sin(\delta_B-\gamma) s_i],
\end{split}
\\
\begin{split}
\hat\Gamma^-_{\bar{i}} &\equiv 
\int_{\bar{i}} d\hat\Gamma(B^- \to (K_S \pi^- \pi^+)_D K^-)=\\
&T_{\bar{i}} +r_B^2 T_i \;+ 
2 r_B [\cos(\delta_B-\gamma) c_i - \sin(\delta_B-\gamma) s_i],
\end{split}
\\
\begin{split}
\hat\Gamma^+_i &\equiv\int_i d\hat\Gamma(B^+ \to (K_S \pi^- \pi^+)_D K^+)=\\
&T_{\bar{i}} +r_B^2 T_i \;+ 
2 r_B [\cos(\delta_B+\gamma) c_i - \sin(\delta_B+\gamma) s_i],
\end{split}
\\
\begin{split}
\hat\Gamma^+_{\bar{i}} &\equiv 
\int_{\bar{i}} d\hat\Gamma(B^+ \to (K_S \pi^- \pi^+)_D K^+)=\\
&T_i +r_B^2 T_{\bar{i}} \;+ 
2 r_B [\cos(\delta_B+\gamma) c_i + \sin(\delta_B+\gamma) s_i].
\end{split}
\end{align}
\end{subequations}
These equations are related to each other through
$12\leftrightarrow 13$ and/or $\gamma \leftrightarrow -\gamma$  exchanges.
All in all,
there are $2k + 3$ unknowns in \eqref{relations4k},
\begin{equation} \label{totpar}
c_i, ~s_i, ~r_B, ~\delta_B, ~\gamma,
\end{equation}
so that the $4k$ relations \eqref{relations4k} are solvable for $k\ge
2$. In other words, a partition of the $D$ meson Dalitz plot to four
or more bins allows for the determination of $\gamma$
without hadronic uncertainties.  This is our main result.

When $c_i=0$ or $s_i=0$ for all $i$, some equations become degenerate
and $\gamma$ cannot be extracted. However, due to resonances, we do
not expect this to be the case. Degeneracy also occurs if
$\delta_B=0$.  In this case, $\gamma$ can still be extracted if some
of the $c_i$ and/or $s_i$ are independently measured, as discussed in the
following sections.

\section{Improved Measurement of \boldmath $c_i$ and $s_i$}
\label{improving}

So far, we have used the $B$ decay sample to obtain all the unknowns,
including $c_i$ and $s_i$, which are parameters of the charm system.
We now show that the $c_i$ and $s_i$ can be independently measured at a charm
factory~\cite{Soffer:1998un,Silva:1999bd,Gronau:2001nr}. This is done
by running the machine at the $\psi(3770)$ resonance, which decays
into a $D\Dbar$ pair. If one $D$ meson is detected in a CP eigenstate
decay mode, it tags the other $D$ as an eigenstate of the opposite CP
eigenvalue. The difference between the two decay widths gives \cite{Giri:2003ty}
\begin{equation}
\begin{split}
c_i = \frac{1}{2} \Big[&\int_i 
	 d\Gamma(D^0_+ \to K_S(p_1) \pi^-(p_2) \pi^+(p_3))\\
	-&\int_id\Gamma(D^0_- \to K_S(p_1) \pi^-(p_2) \pi^+(p_3))
	\Big].
\end{split}
\label{ci-charm-factory}
\end{equation}
where we have defined $D^0_\pm \equiv (D^0 \pm \DbarZ)/\sqrt{2}$.
As stated above, obtaining these independent measurements reduces the 
error in the measurement of $\gamma$ by removing $k$ of the $2k+3$ unknowns.

\begin{figure}
\begin{center}
\epsfig{file=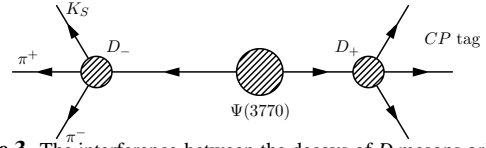, height=1.9cm}
\vspace{-0.5cm}
\caption{\footnotesize{The  interference between the decays of $D$
mesons originating from $\psi(3770)$ allow for a measurement of $c_i$
and $s_i$ at charm factories. Shown is a decay allowing for determination of $c_i$.}} \label{fig-Psidecay}
\end{center}
\vspace{-0.7cm}
\end{figure}

In addition, if one of the $D$ mesons decays into a non-CP
eigenstate, we are sensitive to the $s_i$ variables as well. Consider for
instance a $\psi (3770)$  decaying into a $D,\bar{D}$ pair, of which
one decays into $K_S\pi^+\pi^-$ and the other decays into some
general state $g$. The partial  decay width corresponding to the $i-$th
bin of the $K_S\pi^+\pi^-$ Dalitz  plot and the $j-$th bin of the $g$ 
final state's phase space is
\begin{equation}
\Gamma_{i,j}\propto T_i T_{\bar{j}}^g+T_{\bar i} T_j^g-2(c_i c_j^g+s_i
s_j^g), \label{gen_charm}
\end{equation}
where $T_j^g$, $c_j^g$, $s_j^g$ are defined as in \eqref{defvar}.
In particular, if one chooses $g=K_S\pi^+\pi^-$ and $j=i$ (or
$j=\bar{i}$) one measures $s_i^2$. If, on the other hand,
$g$ is a CP even (odd) eigenstate, $s_j^g=0$, $T_j^g=T_{\bar{j}}^g=\pm
c_j^g$  and equation \eqref{gen_charm}  reduces to
\eqref{ci-charm-factory}.

\section{Assuming Breit-Wigner dependence}\label{Breit-Wigner}
If the functional dependence of both the moduli and the phases of the
$D^0$ meson decay amplitudes $A_D(s_{12},s_{13})$ were known, then the
analysis would be simplified.  There would be only three
variables, $r_B, \delta_B$, and $\gamma$, that need to be fit to the
reduced partial decay widths in Eq.~\eqref{decay-width}. A plausible
assumption about their forms is that a
significant part of the three-body $D^0\to K_S \pi^- \pi^+$ decay
proceeds via resonances. These include decay transitions of the form
$D^0\to K_S \rho^0 \to K_S \pi^- \pi^+$ or $D^0\to K^{*-}(892) \pi^+\to
K_S \pi^- \pi^+$, as well as decays through higher resonances,
e.g., $f_0(980)$, $f_2(1270)$, $f_0(1370)$ or $K^*_0(1430)$. An important feature is that there exists an overlap
region of Cabibbo allowed $D^0\to K^{*-}(892) \pi^+$, $D^0\to K_S
\rho^0$ decays, where the variation of strong phase will be large,
allowing for the extraction of $\gamma$.

The decay amplitude can then be fit to a sum of Breit-Wigner
functions and a constant term. Following the notations of Ref.
\cite{Aitala:2001zx} we write
\begin{equation} 
\begin{split}
A_D(s_{12},s_{13})&=A(D^0 \to K_S(p_1) \pi^-(p_2) \pi^+(p_3))=\\ &=
a_0 e^{i\delta_0}+\sum_r a_r e^{i\delta_r} {\cal A}_r 
(s_{12}, s_{13}),
\end{split}\label{resonansatz}
\end{equation}
where the first term corresponds to the non-resonant term and the second
to the resonant contributions with $r$ denoting a specific
resonance. The functions ${\cal A}_r$ are  products of Breit-Wigner
functions and appropriate Legendre polynomials that account for the fact that $D$
meson is a spin $0$ particle.  
Explicit  expressions can be found in 
Ref. \cite{Aitala:2001zx}.

One of the strong phases $\delta_i$ in the ansatz \eqref{resonansatz} 
can be put to zero, while others are fit to the experimental data 
together with the amplitudes $a_i$.
 The obtained functional form of
$A_D(s_{12},s_{13})$ can then be fed to Eq. \eqref{decay-width}, which
is then fit to the Dalitz plot of the $B^\pm\to (K_S \pi^- \pi^+)_D
K^\pm$ decay with $r_B$, $\delta_B$ and $\gamma$ left as free
parameters. 

\section{Discussions}\label{discussion}
The observables $\hat\Gamma^\pm_i$ defined in 
\eqref{relations4k} can be used to experimentally look for
direct CP violation. Explicitly,
\begin{equation} \label{dircp}
a_{\rm CP}^{i,\bar{i}}\equiv \hat\Gamma^-_{i,\bar{i}} - \hat\Gamma^+_{\bar{i},i} = 
4 r_B \sin\gamma \left[c_i \sin \delta_B \mp s_i  \cos \delta_B\right] ,
\end{equation} 
Nonzero  $a_{\rm CP}$ requires non-vanishing strong
and weak phases.  Due to the resonances, we expect the strong phase to be
large. Therefore, it may be that direct CP
violation can be established in this mode even before the full
analysis to measure $\gamma$ is conducted. With more data, $\gamma$ can be
extracted assuming Breit-Wigner resonances (cf. section
\ref{Breit-Wigner}).  Eventually, a model independent extraction of
$\gamma$ can be done (cf. section~\ref{model-indep} and~\ref{improving}).

The above proposed method for the model independent measurement of
$\gamma$ involves a four-fold ambiguity in the extracted value. The
set of equations
\eqref{relations4k} is invariant under each of the two discrete
transformations
\begin{align}
P_\pi&\equiv \{ \delta_B\to \delta_B + \pi, \gamma\to \gamma +\pi\},\\
P_-&\equiv \{ \delta_B\to -\delta_B, \gamma\to -\gamma,s_i\to -s_i\}. 
\end{align}
The discrete transformation $P_\pi$ is a symmetry of the amplitude
\eqref{amplitude} and is thus an irreducible uncertainty of the
method. It can be lifted if the sign of either $\cos\delta_B$ or
$\sin\delta_B$  is known. The ambiguity due to $P_-$ can be resolved if
the sign of $\sin\delta_B$ is known or if the sign of $s_i$ can be
determined in at least some part of the Dalitz plot for instance by
fitting a part of the Dalitz plot to Breit-Wigner functions. 

The $r_B$ suppression present in the scheme outlined above can be
somewhat lifted if the cascade decay $B^-\to D X_s^- \to (K_S \pi^-
\pi^+)_D X_s^-$ is used \cite{Gronau:2002mu,Aleksan:2002mh}.  Here
$X_s^-$ is a multibody hadronic state with an odd number of kaons
(for instance  $K^-\pi^-\pi^+$, $K^-\pi^0$ or $K_S \pi^-
\pi^0$). The same formalism as outlined above applies also to this case
with trivial changes \cite{Giri:2003ty}. In addition to using different $B$ modes, statistics may be increased by 
employing various $D$ decay modes as well. An interesting possibility
are 
 Cabibbo allowed,  $D\to K_S
\pi^-\pi^+ \pi^0$,~~$K^- K^+ K_S$, and 
Cabibbo suppressed, $D\to K^- K^+ \pi^0$,~~$\pi^- \pi^+ \pi^0$,~~$K_S
K^+ \pi^-$ decay modes, to which  our formalism applies with very minor
changes \cite{Giri:2003ty}.

In conclusion, we have shown that the angle $\gamma$ can be determined
from the cascade decays $B^\pm\to K^\pm (K_S \pi^- \pi^+)_D$.  The
reason for the applicability of the proposed method lies in the
presence of resonances in the three-body $D$ meson decays that provide
a necessary variation of both the phase and the magnitude of the decay
amplitude across the phase space. The fact that no Cabibbo suppressed
$D$ decay amplitudes are used in the analysis is another advantage of
the method and leads to a sensitivity on $\gamma$ at order $O(r_B)$.  However, it does involve a Dalitz plot analysis with
possibly only parts of the Dalitz plot being practically useful for
the extraction of $\gamma$. In reality, many methods have to
be combined in order to achieve the required statistics for a precise
determination of $\gamma$ \cite{Soffer:1999dz}.

\end{document}